\documentclass[graybox]{svmult}

\usepackage{type1cm}        % activate if the above 3 fonts arehttps://www.overleaf.com/project/63c1843ddf3a8f5efdda3a0d
\usepackage{makeidx}         % allows index generation
\usepackage{graphicx}        % standard LaTeX graphics tool
\usepackage{multicol}        % used for the two-column index
\usepackage[bottom]{footmisc}% places footnotes at page bottom

\usepackage{natbib}
\usepackage{newtxtext}       % 
\usepackage[varvw]{newtxmath}       % selects Times Roman as basic font

\usepackage{booktabs}
\usepackage{lscape}
\usepackage{longtable}
\usepackage{multirow}
\usepackage{rotating}
\usepackage{tabularx}
\usepackage{float}
\usepackage{url}
\usepackage{bm}
\usepackage{subfig}

\makeindex             % used for the subject index
\begin{document}

\title*{Functional zoning of biodiversity profiles}
% Use \titlerunning{Short Title} for an abbreviated version of
% your contribution title if the original one is too long
\author{Natalia Golini, Rosaria Ignaccolo, Luigi Ippoliti, Nicola Pronello}
% Use \authorrunning{Short Title} for an abbreviated version of
% your contribution title if the original one is too long
\institute{Natalia Golini \at Department of Economics and Statistics "Cognetti de Martiis", University of Turin, \email{natalia.golini@unito.it}
\and Rosaria Ignaccolo \at Department of Economics and Statistics "Cognetti de Martiis", University of Turin,  \email{rosaria.ignaccolo@unito.it}
\and Luigi Ippoliti \at Department of Economics, University "G. d'Annunzio", \email{luigi.ippoliti@unich.it}
\and Nicola Pronello \at Department of Neuroscience, Imaging and Clinical Sciences, University "G. d'Annunzio", \email{nicola.pronello@unich.it}}
%
% Use the package "url.sty" to avoid
% problems with special characters
% used in your e-mail or web address
%
\maketitle

\abstract*{}

\abstract{Spatial mapping of biodiversity is crucial to investigate spatial variations in natural communities. Several indices have been proposed in the literature to represent biodiversity as a single statistic. 
However, these indices only provide information on individual dimensions of biodiversity, thus failing to grasp its complexity comprehensively. Consequently, relying solely on these single indices can lead to misleading conclusions about the actual state of biodiversity.
In this work, we focus on \textit{biodiversity profiles}, which provide a more flexible framework to express biodiversity through non-negative and convex curves, which can be analyzed by means of functional data analysis. By treating the whole curves as single entities, we propose to achieve a \textit{functional zoning} of the region of interest by means of a penalized model-based clustering procedure. 
This provides a spatial clustering of the biodiversity profiles, which is useful for policy-makers both for conserving and managing natural resources and revealing patterns of interest. Our approach is discussed through the analysis of \textit{Harvard Forest Data}, which provides information on the spatial distribution of woody stems within a plot of the Harvard Forest.}


%\jnlcitation{\cname{%
%\author{N. Golini}, 
%\author{R. Ignaccolo}, 
%\author{L. Ippoliti}, and 
%\author{N. Pronello}} (\cyear{20??}), 
%\ctitle{Functional zoning of biodiversity profiles}, \cjournal{Environmetrics}, \cvol{20??;??:??--??}.}

%\footnotetext{\textbf{Abbreviations:} ANA, anti-nuclear antibodies; APC, antigen-presenting cells; IRF, interferon regulatory factor}

\section{Introduction}\label{sec1:intro}

Biodiversity, or biological diversity, is the scientific term indicating the variability among all living organisms in a given area and representing a general indicator of the overall ecological health (e.g. human health and well-being, animal and environmental health, see \cite{delong1996defining}. Biodiversity is part of applied ecology and encloses the diversity within species, the diversity between species and the diversity of ecosystems. The human species, through its actions and activities, has played a significant role in contributing to the biodiversity loss that we can observe today.
Obviously, a biodiversity decline implies a decline in populations, genes, and ecosystems. All these are the irreversible consequences of environmental change affecting human health and well-being \citep{diaz2006biodiversity,cardinale2012biodiversity, Schmeller2020}. To stop this harmful chain, many organizations, agencies, and commissions have established expert working groups or initiatives to monitor, protect and restore biodiversity (see \citealp{DIAZ20151,WHO:2020,EU:2021,FAO:2022} among others). At the basis of these actions, a quantitative measurement of the complex concept of biodiversity is essential, as well as its spatial and temporal change.

In literature, many mathematical functions, called \textit{biodiversity indices}, have been proposed \citep{MAGURRAN2021,Pielou1975}. Each proposed index measures biodiversity from a different perspective, reflecting researchers' various interests in measuring biodiversity (e.g. counting the number of species present in a given area or describing the compositional change of the species abundance distribution). 
As a result, there is currently no consensus on which indices provide a more accurate measure of biodiversity.

In this work, we consider the \textit{species/taxonomic} diversity in the Hill numbers framework based on the notion of \textit{effective number of species} \citep{Hill73,ChaoColwell2022}. The Hill numbers refer to a family of species diversity indices defined for a parameter $q \in [0, +\infty)\symbol{92} \{1\}$, called \textit{order} of the diversity, that gives information about the species abundance distribution. Mathematically, they can be represented as a positive, decreasing, and convex curve of the parameter $q$. 
Then, the \textit{biodiversity profiles}, or \textit{curves}, can be regarded as constrained functional data and, therefore, can be analyzed using functional data analysis (FDA) techniques \citep{ramsay2005functional,ferraty2006nonparametric}. 
A functional approach to biodiversity profiles was initially proposed by \cite{gattone2009functional}, who used a functional linear regression model to assess the impact of habitat effects on diversity changes. Our focus, instead, shifts towards clustering functional data indexed by the cells of a finite spatial lattice, aiming to promote a concept known as \textit{functional zoning} of biodiversity profiles. This approach combines functional data analysis with spatial clustering techniques, identifying homogeneous zones which may serve as a valuable tool for policymakers, enabling them to effectively conserve and manage natural resources while revealing significant patterns of interest.

Although functional data analysis has gained significant attention across various research fields, there has been relatively limited progress in the domain of functional data clustering, especially when considering spatially dependent functions - see, for example, the discussion in the recent review by \cite{reviewFunClust2023}.

Proposals in the frameworks of hierarchical and dynamic clustering approaches, where the similarity between pairs of curves is based on the use of the variogram function, are given by \cite{giraldo2012hierarchical}, \cite{romano2015performance} and \cite{romano2017spatial}. Other approaches based on the use of spatial heterogeneity measures and spatial partitioning methods were also proposed by \cite{Dabo2010}, \cite{secchi2013voronoi} and  \cite{fortuna2020unsupervised}. A few proposals can also be found in the framework of model-based approaches. \cite{Vandewalle2020} and \cite{WuLi2022}, for example,  incorporate longitude and latitude coordinates as regressors in a multinomial logistic regression model, which is employed to estimate the prior probabilities of a mixture model. On the other hand, \cite{Jiang2012} and \cite{LiangETal2021} utilize Markov Random Fields and Gibbs distribution to account for spatial dependence in their clustering procedures.

In this paper, we also use a model-based approach for spatially correlated functional data. In particular, we consider a penalized model-based clustering procedure where a finite mixture of Gaussian distributions is used to model the expansion coefficients obtained from approximating the functional biodiversity profiles in a finite-dimensional space. To take care of the presence of spatial correlation, the procedure allows the modelling of the spatial distribution of the weights of the mixture such that observations corresponding to nearby locations are more likely to have similar allocation probabilities than observations that are far apart in space. The procedure represents a generalisation of the approach proposed in \cite{Vandewalle2020}, and implementation details are provided in \cite{pronello2022}. 
In the following, we show that this approach proves to be useful for achieving a \textit{functional zoning} of biodiversity profiles in the context of the Harvard Forest Data, a well-known collection of datasets \citep{harvard2022data} that includes two censuses of all woody stems with a minimum diameter of $1cm$ at breast height. We note that the dataset referring to the first census was previously analyzed by \cite{fortuna2020unsupervised}. However, in their analysis, they performed an exploratory analysis and identified spatial outliers before obtaining spatial clustering only on a limited number of diversity profiles. They achieved this through the use of a distance-based LISA map in both hierarchical and k-means algorithms. 

The paper is structured as follows. In Section \ref{sec2:data} we provide a brief description of the motivating example and the data used in this study. In Section \ref{sec3:biodiv} we summarize the key conceptual issues underlying the measurement of biodiversity, discuss some of the most commonly used diversity indices, their conversion to effective numbers and the derivation of biodiversity profiles. Section \ref{sec4:FDA} introduces the functional representation of biodiversity profiles and proposes empirical variogram functions to characterize their possible spatial dependence structure. Section \ref{sec5:GMM} presents the finite Gaussian Mixture Model (GMM) employed for spatial clustering purposes, while Section \ref{sec6:results} illustrates the results of the functional zoning of biodiversity profiles for the Harvard Forest Data. Lastly, Section \ref{sec7:discussion} concludes the paper by offering conclusions and suggestions for future research.

\section{The motivating case study}\label{sec2:data}

Forests play a crucial role in tackling biodiversity conservation and restoration. According to \cite{FAO2020UNEP}, forests cover almost one-third of the global land area and harbour most of the terrestrial biodiversity. So it is essential to provide policymakers with a tool to prioritize forestry policies and implement plans that positively impact biodiversity at the population, genetic and ecosystem levels.

\begin{figure}[t]
    \centering
    \includegraphics[scale = 0.25]{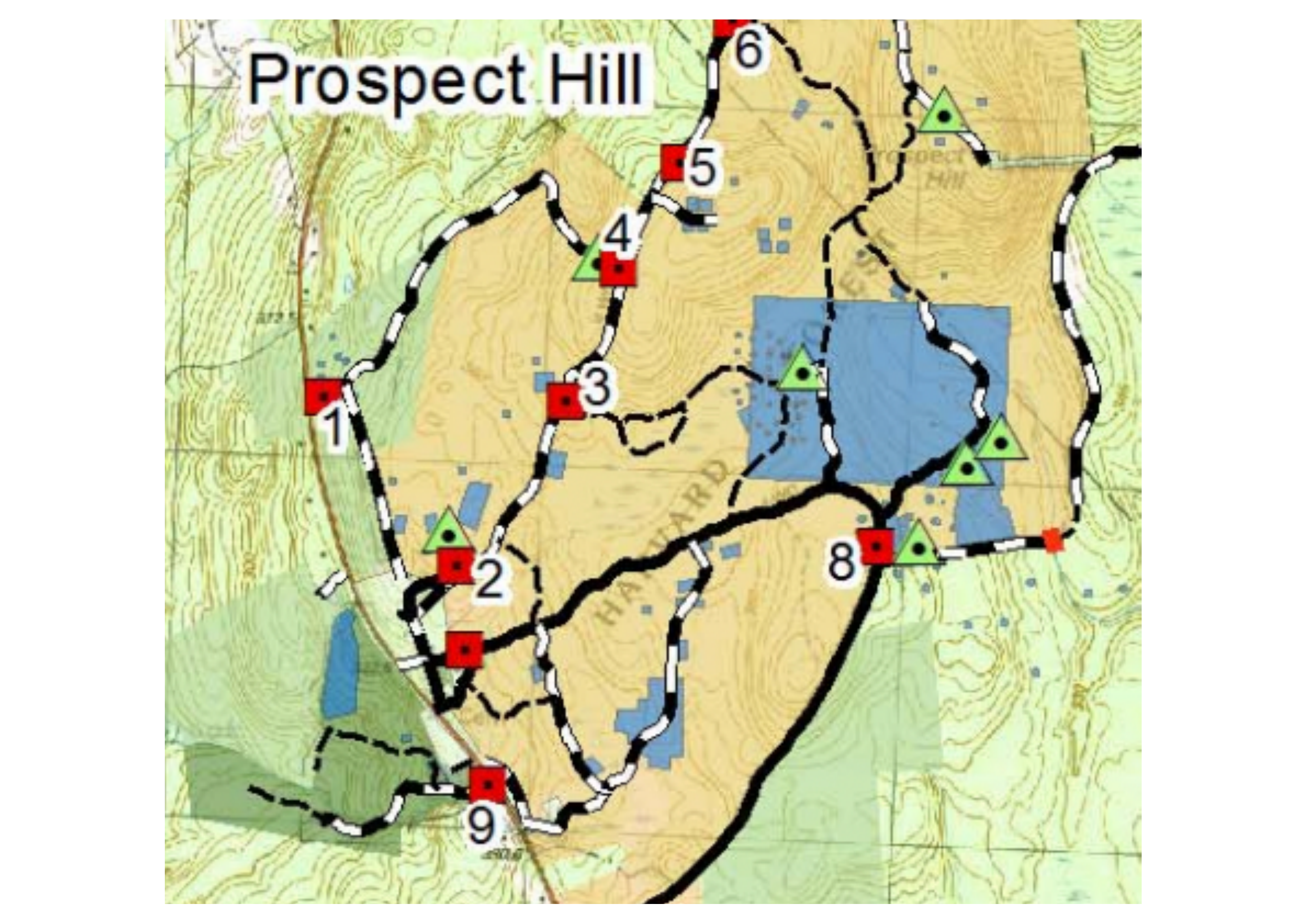}
    \caption{The Prospect Hill Tract long-term plot. The blue rectangle area represents the long-term study area of interest.} \label{HFmap}
\end{figure}

Harvard Forest is a vast laboratory and classroom of Harvard University, where observational studies and experiments are conducted 
to drive research and education on several topics. One of the most relevant is the study of biodiversity.
Harvard Forest provides detailed inventories of species diversity. An example of a dataset (data and metadata) for biodiversity studies is the  \textit{Harvard Forest CTFS-ForestGEO Mapped Forest Plot since 2014} (Number ID HF253, version 5, \citealp{harvard2022data}), where data were collected within the $35ha$ plot located on Prospect Hill (see Figure \ref{HFmap}), the hub of research activity performed at Harvard Forest (Petersham, Massechussen, New England region). This plot is one of the seventy-four Center for Tropical Forest Science-Forest Global Earth Observatory (CTFS-ForestGEO).\footnote{CTFS-ForestGEO is a worldwide network monitoring forests for advancing the long-term study of forest dynamics and biodiversity. See \url{https://forestgeo.si.edu/} for more details.} It covers a rectangle area of size $500 m \times 700 m$, and it was designed to \textit{include a continuous, expansive, and varied natural forest landscape} \citep{harvard2022data}, and it is a continuous grid of $875$ cells of size $20m \times 20 m$.

\begin{figure}[t]
    \centering
    \includegraphics[scale = 0.35]{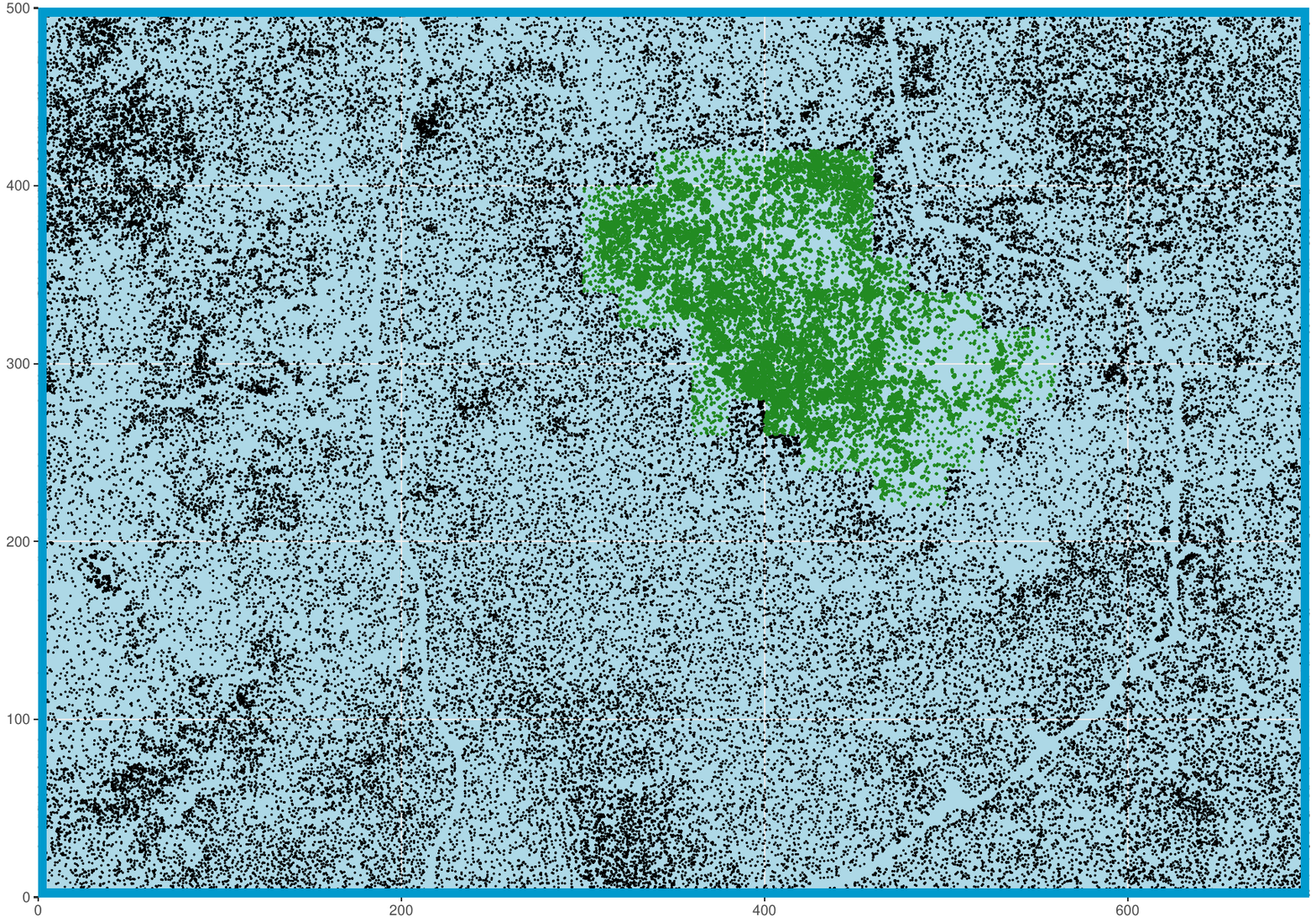}
    \caption{Distribution of the woody stems greater than $1cm$ diameter at breast height collected within the Prospect Hill Tract long-term plot ($500 m \times 700 m$). In black are the data collected during the second census (May 2018 - January 2020); in green are the data collected during the first census (June 2010 - March 2014) in the swamp area.} \label{HFmap_integrated}
\end{figure}

HF253 is a collection of five datasets freely available for download at \url{https://harvardforest1.fas.harvard.edu/exist/apps/datasets/showData.html?id=HF253}. 
In particular, we are interested in the most recent dataset "hf253-05”, consisting of $85,641$ woody stems greater than $1cm$ diameter at $1.3m$ (at breast height) collected between May 2018 and January 2020 (second census). However, this census does not contain data from the swamp in the plot's central portion. Data collection in this area was supposed to take place during the winter of 2021 but was not carried out due to restrictions related to the COVID pandemic. Moreover, a winter census for the swamp area was not planned for 2022. 
Given the unique characteristics of the swamp area, we made the decision not to impute the missing data in this region by means of a statistical technique. Instead, we replaced the missing values with the $37,577$ observations collected for the swamp area during the first census, which took place from June 2010 to March 2014.
Figure \ref{HFmap_integrated} shows the available data within the Prospect Hill Tract long-term plot. In black are displayed the data collected during the second census, while in green we show the data collected during the first census in the swamp area. 
Then, the complete dataset consists of $123,218$ records providing information on each collected stem, identified by a unique identifier (\texttt{stem.id}) representing the primary key of the dataset. 
However, only some information is of interest for our analysis, specifically: the species mnemonic (the full Latin name, the family and other information on the species are available in the dataset "hf253-02"), the coordinates in meters ($m$) within the plot relative to the left-down corner of the area of interest, the diameter of the stem in centimetres ($cm$) and the status of the stem (alive, dead, lost stem, missing, prior). It is crucial to emphasize here that the terms "alive" and "dead" refer to the whole tree. If any stem remains alive, the tree is considered alive. The tree is deemed dead only when every single stem has perished.
Given this information, we can calculate abundance data for each tree species within each of the 875 cells of the grid covering the Prospect Hill Tract long-term plot. 

In this application, we first perform a pre-processing step to focus on the stems that possess the "alive" status and have a diameter exceeding five cm, obtaining 34,287 woody stems. 
To retrieve the trees, we filtered the pre-processed stems dataset for unique rows based on the tree identifier (\texttt{tree.id}). This process resulted in a total of 31,153 individual trees, representing 37 different species that are mapped over the area of interest. Of these $31,153$ trees, only $3,140$ have more than one stem. Table \ref{species_counts} presents the species-wise distribution of tree abundances. The most frequently occurring species in the Prospect Hill Tract long-term plot are listed in the first seven positions of Table \ref{species_counts}. Among these species, \textit{Tsuga canadensis} and \textit{Acer rubrum} can be considered dominant. However, it is important to note that many rare species are also present in the Prospect Hill Tract long-term plot, indicating the presence of biodiversity.

\begin{table}[t]%
\caption{Absolute abundances of trees grouped by species. The species mnemonic and the full Latin name are reported for each species. } \label{species_counts}
\centering
\begin{tabular}{llr|llr}
\toprule
\textbf{Species mnemonic} & \textbf{Full genus and species name}  & \textbf{Count}  & \textbf{Species mnemonic} & \textbf{Full genus and species name}  & \textbf{Count}   \\
\midrule
tsugca & \textit{Tsuga canadensis} & 11673 & betupo & \textit{Betula populifolia} &   30\\
acerru & \textit{Acer rubrum} & 7364 &  queral & \textit{Quercus alba} &   25\\
querru & \textit{Quercus rubra} & 3388 &  alnuin & \textit{Alnus incana} &   21\\
betual & \textit{Betula alleghaniensis} & 2342 &  amella & \textit{Amelanchier laevis} &   21\\
pinust & Pinus strobus & 1395 &  fraxni & \textit{Fraxinus nigra} &   15\\
fagugr & \textit{Fagus grandifolia} & 1352 &   sorbam & \textit{Sorbus americana} &   15\\
betule & \textit{Betula lenta} & 948 &  ostrvi  & \textit{Ostrya virginiana} &  12\\
pinure & \textit{Pinus resinosa} & 547 &  picexx  & \textit{Picea unknown} &  12\\
hamavi &\textit{Hamamelis virginiana} & 343 & ilexve   & \textit{Ilex verticillata} & 10\\
kalmla & \textit{Kalmia latifolia} & 319 &  querxx  & \textit{Quercus unknown} &  10\\
betupa & \textit{Betula papyrifera} & 262 &  vaccco  & \textit{Vaccinium corymbosum} &   9\\
piceab & \textit{Picea abies} & 236 &  nemomu  & \textit{Nemopanthus mucronatus} &   7\\
querve & \textit{Quercus velutina} & 181 &  toxive  & \textit{Toxicodendron vernix} &   5\\
nysssy & \textit{Nyssa sylvatica} & 136 &  betuxx  & \textit{Betula unknown} &   4\\
prunse & \textit{Prunus serotina} & 120 &  popugr  & \textit{Populus grandidentata} &   2\\
castde & \textit{Castanea dentata} & 117 &  acersa  & \textit{Acer saccharum} &   1\\
fraxam & \textit{Fraxinus americana} & 101 &  ilexmu  & - &   1\\
acerpe & \textit{Acer pennsylvanicum} & 65 &  pinuxx   & \textit{Pinus unknown} &  1\\
piceru & \textit{Picea rubens} & 63 & & & \\
\bottomrule
\end{tabular}
\end{table}

\begin{figure}[t]
    \centering
    \includegraphics[scale = 0.5]{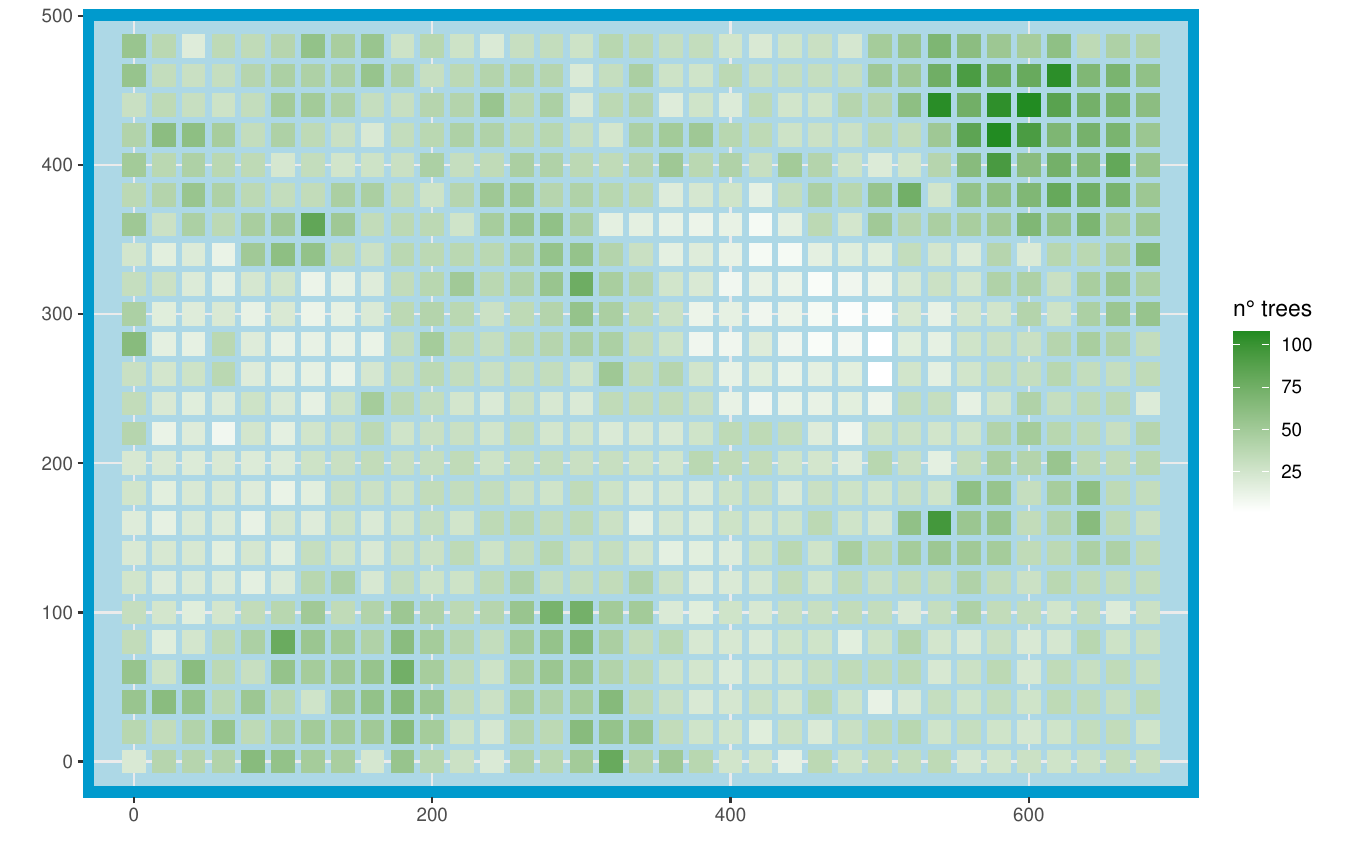}
    \caption{Absolute number of trees in each of the 875 cells of the Prospect Hill Tract long-term plot.}\label{nCounts}
\end{figure}

\begin{figure}[t]
    \centering
    \includegraphics[scale = 0.5]{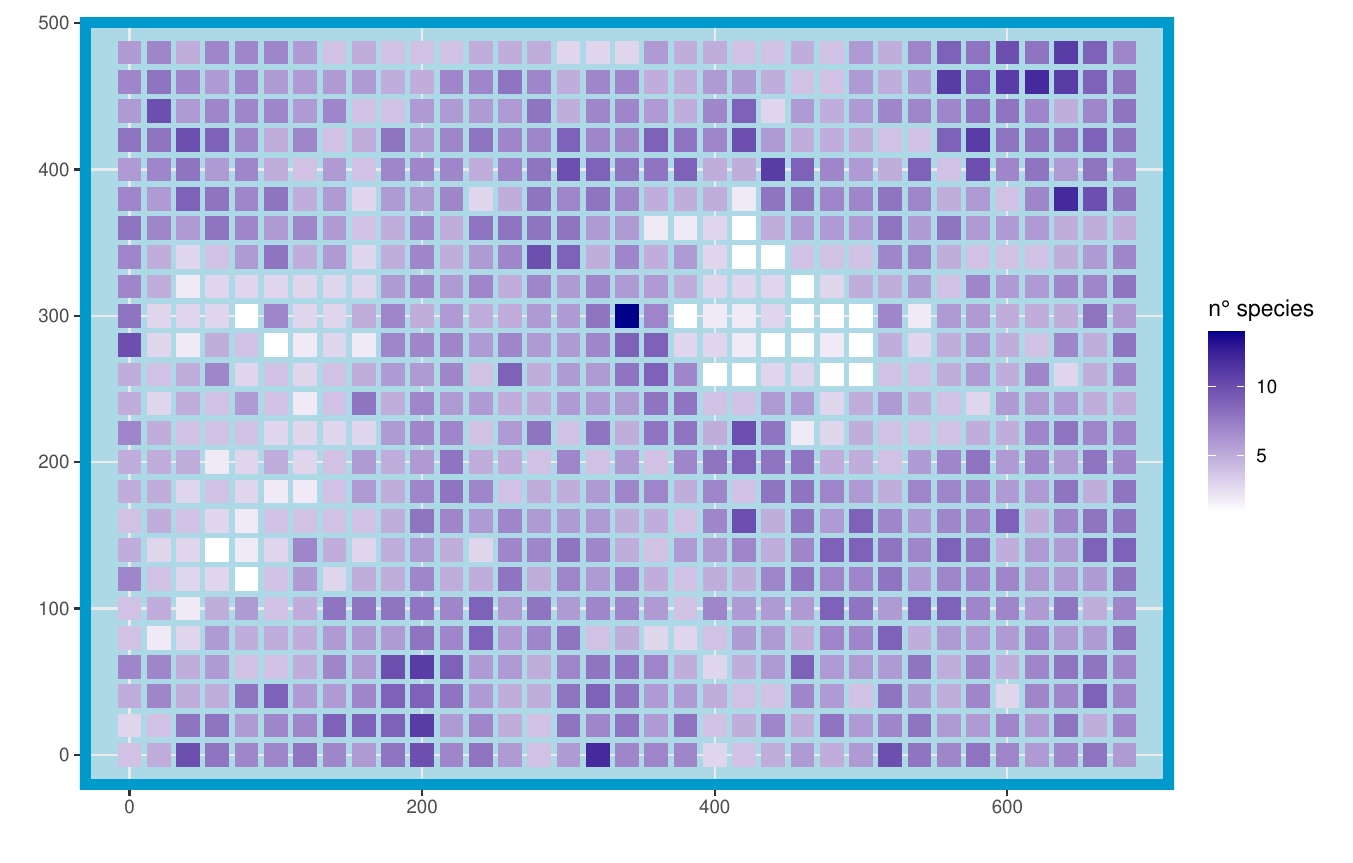}
    \caption{Absolute number of species in each of the 875 cells of the Prospect Hill Tract long-term plot (species richness).}\label{nSpecies}
\end{figure}

\begin{figure}[t]
         \centering
         \includegraphics[scale = 0.6]{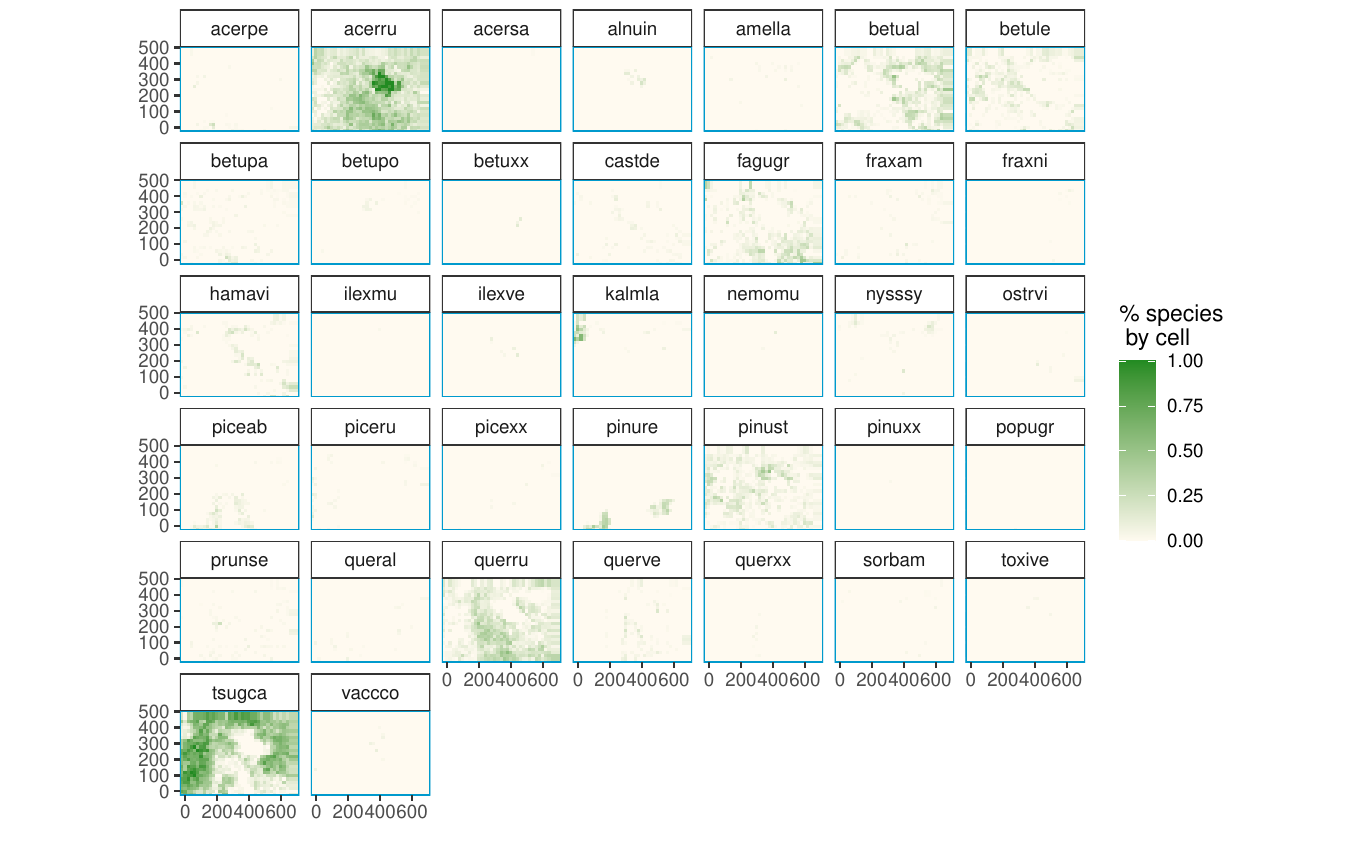}
        \caption{Spatial distribution of the relative abundance of species in each of the 875 cells of the Prospect Hill Tract long-term plot (species evenness). }\label{percSpecies}
\end{figure}

Figure \ref{nCounts} shows the absolute number of trees detected in each Prospect Hill Tract long-term plot cell. The most populated area is the one relative to the right-up corner of the Prospect Hill Tract long-term plot. In this area, it is possible also to note the higher species richness, i.e. the absolute number of species present in each cell of the Prospect Hill Tract long-term plot (see Figure \ref{nSpecies}). Figure \ref{percSpecies} shows that  \textit{Tsuga canadensis}, \textit{Acer rubrum}, and \textit{Betula alleghaniensis} are, among the other species, more present in this area. %\sout{(species evenness)}. 
This information provides evidence of species evenness, i.e. in a cell the community is perfectly even if every species is present in equal proportions and uneven if one species is dominant.
The swamp area records a few trees belonging to the same species, the \textit{Acer rubrum} (acerru) - see Figures \ref{nCounts}, \ref{nSpecies} and \ref{percSpecies}.

The descriptive analysis conducted on the Prospect Hill Tract long-term plot yields valuable insights into various aspects of biodiversity. It offers information on species richness, evenness, and the dominance of specific species, which are important indicators of biodiversity. However, it is important to note that no single measure can fully capture the complexity and entirety of biodiversity within this ecosystem. Biodiversity is a multifaceted concept that extends beyond solely considering the number and distribution of species. In the next section, we will thus delve into the challenge of measuring biodiversity and consider the use of biodiversity profiles as a method to address this complex issue. 

\section{Measuring biodiversity}\label{sec3:biodiv}

In conservation ecology, information on the spatial distribution and composition of biological communities is essential in designing effective biodiversity conservation and management strategies. Site clustering and prioritization are crucial because resources for conservation are often limited, and it is essential to allocate them effectively to maximize conservation outcomes. 

Biodiversity, primarily considered here as \textit{taxonomic diversity}, can be measured in various ways, depending on the study's specific objectives. Common measures of biodiversity include solely species richness or species evenness alone. However, biodiversity is a complex and multivariate concept, and attempting to measure it using a single index has its limitations. While such indices offer valuable insights into specific aspects of biodiversity, they often fail to capture the full richness and intricacies of this multifaceted phenomenon.  

Complexity and multivariate measures have been developed to encompass multiple biodiversity components simultaneously, incorporating information on species composition, abundance, and other ecological attributes. Diversity indices based on species abundance distributions, such as Shannon entropy and Gini-Simpson diversity index, provide a single measure of diversity that considers both richness and evenness. Shannon entropy \citep{Shannon1948} measures the information content or uncertainty associated with the species composition within a community while the Gini-Simpson index \citep{gini1912variabilita,simpson1949measurement} represents the probability that two individuals randomly selected from a community belong to the same species and is the complement of Simpson's original formulation. 
However, interpreting and comparing complex indices can be challenging due to variations in their measurement units and potential non-linear formulations. Shannon entropy is measured in information units, while the Gini-Simpson index is a probability. But more importantly, these indices do not fulfil the \textit{doubling propriety}, an essential requirement for the diversity measures. 
This propriety states that if two communities have equal diversity (measured using certain indices) and an equal number of individuals but do not share any species in common, then the diversity of the pooled community will be twice the diversity of either individual community.

To solve this problem, \cite{macarthur1965patterns} proposed to convert the complexity measures to the \textit{effective number of species}, that is the hypothetical number of equally abundant species that would produce the same value of a diversity measure as the observed community. By converting diversity measures into the effective number of species, researchers can quantify and compare diversity levels more accurately, accounting for differences in species richness and evenness. This approach helps to capture the underlying complexity of biodiversity and provides a more intuitive way to understand and interpret diversity values. For instance, if a diversity measure such as the Shannon entropy or Gini-Simpson index is calculated for a community (e.g. in a cell  of the Prospect Hill Tract long-term
plot), the effective number of species can be derived by transforming the diversity measure into an equivalent number of equally abundant species. Mathematically, Shannon entropy is transformed into its exponential form, and the Gini-Simpson index is converted to the inverse of its complement to $1$ \citep{Jost2006}. 

\subsection{Hill numbers and biodiversity profiles}\label{sub_sec2:1}

The family of the Hill numbers is a family of diversity indices based on the concept of \textit{effective number of species} that allows capturing both species richness and the evenness of species abundances within a community (cell). Hill numbers are expressed as a function of a parameter $q$, which determines the order of the Hill number \citep{Hill73}. 

Given the $N = 875$ cells of the Prospect Hill Tract long-term plot, we assume that each cell contains $S_i, \ \ i=1,\ldots, N$, species of trees. In the following, we denote with $\bm{v}_i$ the $i$-th cell with the spatial coordinates $(x_i,y_i)$ and with $\mathbf{p}_i = \mathbf{p}(\bm{v}_i)=\bigl(p_{1}(\bm{v}_i), \ldots, p_{s}(\bm{v}_i), \ldots, p_{S_i}(\bm{v}_i) \bigr)$ the cell-specific relative abundance vector of species, where $0 \le p_{s}(\bm{v}_i) \le 1$ and $\sum_{s=1}^{S_i} p_{s}(\bm{v}_i)=1$. Then, the family of the Hill numbers is given by

\begin{equation}\label{defHill}
    H(q; \mathbf{p}_i) = \left( \sum_{s=1}^{S_i} p_{s}(\bm{v}_i)^q \right)^{1/(1-q)} \quad \text{for} \quad q \in [0, +\infty)\symbol{92} \{1\} \quad \text{and} \quad i=1, \ldots, N.
\end{equation}
The order $q$ of the Hill number determines the weight given to rare versus abundant species in the diversity evaluation. When $q = 0$, the Hill number represents the species richness. For $q = 1$ the Hill number is not defined, but the limit exists and gives the exponential of the Shannon entropy. When $q = 2$ the Hill number coincides with the inverse of the complement of the Gini-Simpson index. For all $q \ge 0$, Hill numbers satisfy the \textit{doubly property} and have the same measurement unit as species richness. 

To visualize the information captured by Hill numbers across different orders, a \textit{biodiversity profile} can be created by plotting the Hill numbers on a single graph as a function of the parameter $q$. This profile shows how the Hill numbers change as the parameter $q$ varies, providing a comprehensive view of diversity patterns and capturing the multivariate nature of biodiversity. In particular, the region of a biodiversity profile with small values of $q$ provides insights into species richness and rare species since $H(q; \mathbf{p}_i)$ is influenced significantly by both common and rare species. Conversely, the tail of the biodiversity profile with large values of $q$ sheds light on dominance and common species, as $H(q; \mathbf{p}_i)$ becomes less affected by rare species. The order parameter $q$ represents, therefore, the \textit{insensitivity} to rare species. As it grows, the perceived diversity $H(q; \mathbf{p}_i)$ drops.

\begin{figure}[t]
    \centering
    \includegraphics[scale = 0.6]{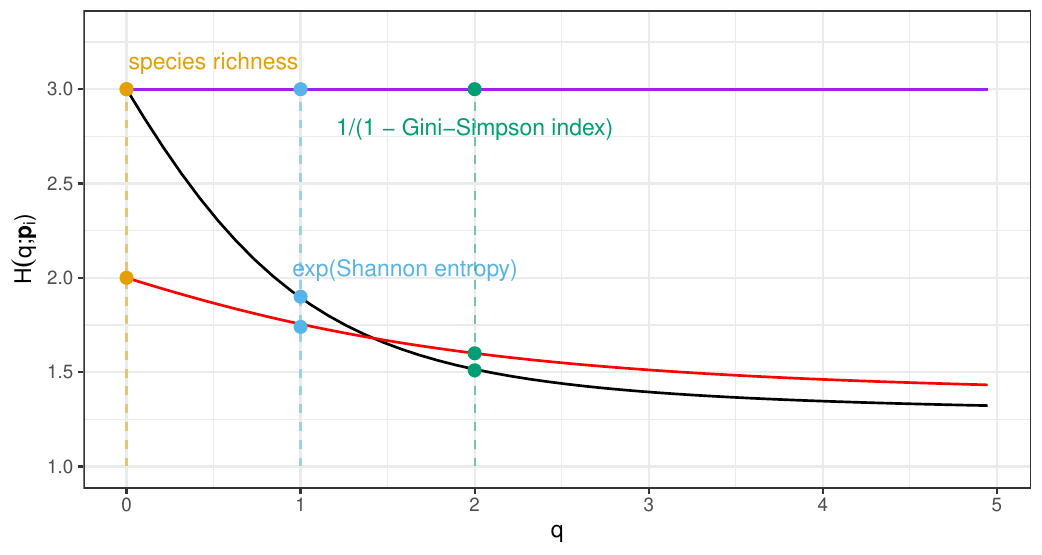}
    \caption{Comparison of three biodiversity profiles considering the parametric family of Hill numbers for two cell-specific relative abundance vectors. In black the biodiversity profile for $\mathbf{p}_1 = \mathbf{p}(\bm{v}_1)=(0.8, 0.1, 0.1)$, in purple the biodiversity profile for $\mathbf{p}_2 = \mathbf{p}(\bm{v}_2)=(0.333, 0.333, 0.333)$ and in red the biodiversity profile for $\mathbf{p}_3 = \mathbf{p}(\bm{v}_3)=(0.75,0.25)$. The Hill number of order $0$ corresponds to species richness, the Hill number of order $1$ is equal to the exponential of the Shannon index, and the Hill number of order $2$ coincides with the inverse of the complement of the Gini-Simpson index. } \label{Hn:exe}
\end{figure}

To better understand the mathematical relationships between the species richness, Shannon entropy, Gini-Simpson index, and Hill numbers, consider the following example. Suppose we have three cells, $\bm{v}_1$,  $\bm{v}_2$ and $\bm{v}_3$, equipped with the following relative abundance vectors: $\mathbf{p}_1 = \mathbf{p}(\bm{v}_1)=(0.8, 0.1, 0.1)$, $\mathbf{p}_2 = \mathbf{p}(\bm{v}_2)=(0.333, 0.333, 0.333)$, and $\mathbf{p}_3 = \mathbf{p}(\bm{v}_3)=(0.75,0.25)$,  whose Hill biodiversity profiles are represented in Figure \ref{Hn:exe}. Individually, the three curves display typical properties of the biodiversity profiles. For example, an ecologist most concerned with species richness would say that the black and purple profiles show three species in the two cells $\bm{v}_1$ and $\bm{v}_2$ and that cell $\bm{v}_3$ (with red profile) has one species less than the others. Furthermore, if one is principally concerned with dominance, it can be noticed that the biodiversity profile for $\bm{v}_2$ is constantly above the others, suggesting that the community in this cell is perfectly even and that it shows the most diverse community type.  On the other hand, the biodiversity profile for $\bm{v}_1$ tends to drop sharply between $q=0$ and $q=2$, levelling off soon after $q = 3$. In particular, the abrupt drop in the region $0 \leq q \leq 1$ indicates that this community has lower biodiversity with more rare species compared with the $\bm{v}_2$ one. In general, as $q$ increases, these rare species are given less weight by the index, and therefore the steeper the drop of the profile, the more rare species there are in the community. Finally, it is possible to note that the black profile crosses the red one at nearly $q=1.5$, suggesting that the community in $\bm{v}_1$ is richer but also moderately more even than that in the cell $\bm{v}_3$. Overall, when two biodiversity profiles cross, the relative rankings of the two profiles depend on the specific diversity order being considered. In other words, their ordering or ranking can only be determined within the context of a specific order parameter $q$. 

\section{Functional Data Analisys for Hill numbers profiles}\label{sec4:FDA}

Let $\mathbf{p}_i = \mathbf{p}(\bm{v}_i)=\bigl(p_{1}(\bm{v}_i), \ldots, p_{s}(\bm{v}_i), \ldots, p_{S_i}(\bm{v}_i) \bigr), \ i=1, \ldots, N$, denote the cell-specific relative abundance vector for $S_i$ species and let $H(q; \mathbf{p}_i)$ be the corresponding biodiversity profile. These profiles can be perceived as samples of (spatially dependent) smooth curves which, in turn, can be viewed as realizations of an underlying biological process generating the abundance vectors  $\mathbf{p}_i$. 

\begin{figure}
         \centering
         \includegraphics[scale = 0.6]{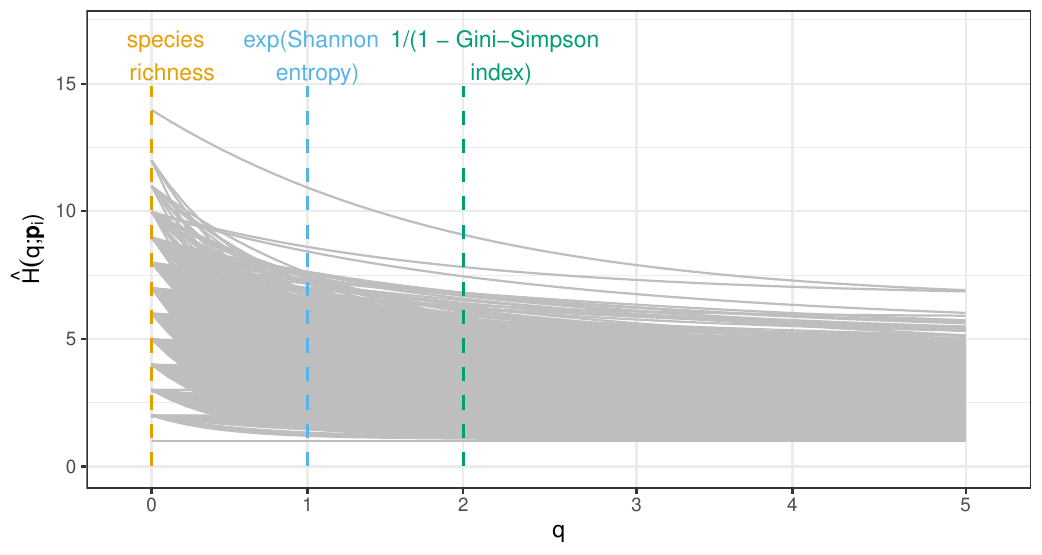}
         \caption{Fitted curves, one per each cell in the Prospect Hill Tract long-term plot.}\label{fitted:curves}
\end{figure}

Following \cite{gattone2009functional}, the biodiversity profiles, $H(q; \mathbf{p}_i)$,
can thus be studied within the FDA framework. 
However, modelling biodiversity profiles is not so straightforward as they are non-negative, monotone decreasing and convex functions over their domain. 
To avoid undesirable effects from their modelling, we make use of the solution proposed by \cite{Ramsay1998}, which was also adopted in the work by \cite{gattone2009functional}. This solution involves representing the function $H$ as a transformation of an unconstrained Lebesgue square integrable function, denoted henceforth as $\tilde{H}$. For each cell, the function $H$
can thus be seen as a solution of the differential equation $D^2 H = \tilde{H} DH$, and it can be written as 
\begin{equation}
H(q; \mathbf{p}_i) = \xi_{0i} + \xi_{1i} \ D^{-2} \Bigl[ \exp\left( D^{-1} \tilde{H}(q; \mathbf{p}_i) \right) \Bigr], \qquad i=1, \ldots, N, %\\
\end{equation}
where $\xi_{0i}$ and $\xi_{1i}$ are arbitrary constants, while $D^m$ and  $D^{-m}$ are the partial differential and integration operators of order $m$, respectively. Being unconstrained, $\tilde{H}$ can be expanded as a linear combination of a finite set of basis functions  $\phi_j(q)$, $j=1, \ldots, J$, so that $$\tilde{H}(q; \mathbf{p}_i) =\sum_{j=1}^J \alpha_{ji} \phi_j(q) $$
and each function $H(q; \mathbf{p}_i)$ can be represented by its vector of coefficients collected in the vector $\bm{\beta}_i = (\xi_{0i}, \xi_{1i}, \alpha_{1i}, \ldots, \alpha_{Ji} )^T, \ i=1, \ldots, N$. By using a penalized regression for each profile, the fitted function takes the form 
\begin{equation}\label{eq:fit_fuct}
\hat{H}(q; \mathbf{p}_i) =
 \hat{\xi}_{0i} + \hat{\xi}_{1i} \ D^{-2} \Bigl[\exp\Bigl( D^{-1} \sum_{j=1}^J \hat{\alpha}_{ji} \phi_j(q) \Bigr)\Bigr], \qquad i=1, \ldots, N. 
\end{equation}
Figure~\ref{fitted:curves} shows all the $875$ fitted curves on Hill number profiles, one per each cell in the Prospect Hill Tract long-term plot, estimated with $J=15$ basis functions and the domain for $q$ truncated at $Q=5$. 
As needed, all the fitted curves are monotone decreasing, and they start from the maximum at $q=0$, which coincides with the species richness. The $875$ curves also cross each other so that the answer to the question "\textit{where is the  Prospect Hill Tract long-term plot most diverse?}" depends heavily on the order parameter $q$: the ranking of the cells may change several times and, to highlight possible similarities in the shape of the whole profiles, in the following we propose a suitable clustering procedure. 

\subsection{Assessing spatial dependence for  functional data}

Standard statistical techniques for modelling functional data primarily focus on independent functions. However, assuming independence appears unreasonable when observing samples of functions across different contiguous cells. Accordingly, when clustering biodiversity profiles in space, it is crucial to assess spatial dependence to understand the underlying spatial patterns and ensure the validity of the clustering results.

Analyzing the spatial variability of biodiversity profiles can be done using a trace-variogram for functions \citep{GiraldoETal2011} defined as 
\begin{equation}
    2{\gamma}(\mathbf{h})=  E \left[
    \int_{0}^{Q} \Bigl({H}\bigl(q; \mathbf{p}_i(\bm{v}_i)\bigr)-{{H}}\bigl(q; \mathbf{p}_i(\bm{v}_i+\mathbf{h}\bigr)\Bigr)^2 dq \right]
    \label{variogram:function}
\end{equation}
over a vector distance $\mathbf{h}$. An important assumption underlying the use of the $L_2$ distance in the trace-variogram
in Eq. (\ref{variogram:function}) is that the length of the domain of the functions is fixed. 
Specifically, the latter assumption assumes perfect alignment of the functions, which is not a concern within the framework of biodiversity profiles.

Under stationarity hypothesis, it is common practice to estimate the trace-variogram in Eq. \ref{variogram:function}) by a mean value of samples grouped over an isotropic distance $h$:
\begin{equation}
    2\hat{\gamma}(h)= \frac{1}{n(h)}\sum_{||\bm{v}_i-\bm{v}_r||=h} 
    \int_{0}^{Q} \bigl({\hat{H}}(q; \mathbf{p}_i)-{\hat{H}}(q; \mathbf{p}_r)\bigr)^2 dq,
    \label{trace_variogram:HP}
\end{equation}
where $n(h)$ is the number of pairs $\bigl(\mathbf{p}(\bm{v}_i),\mathbf{p}(\bm{v}_r)\bigr)$ at spatial distance $h$ and $\hat{H}(\cdot)$ are as defined in Eq.~(\ref{eq:fit_fuct}).

\begin{figure}
    \centering
    \includegraphics[scale = 0.6]{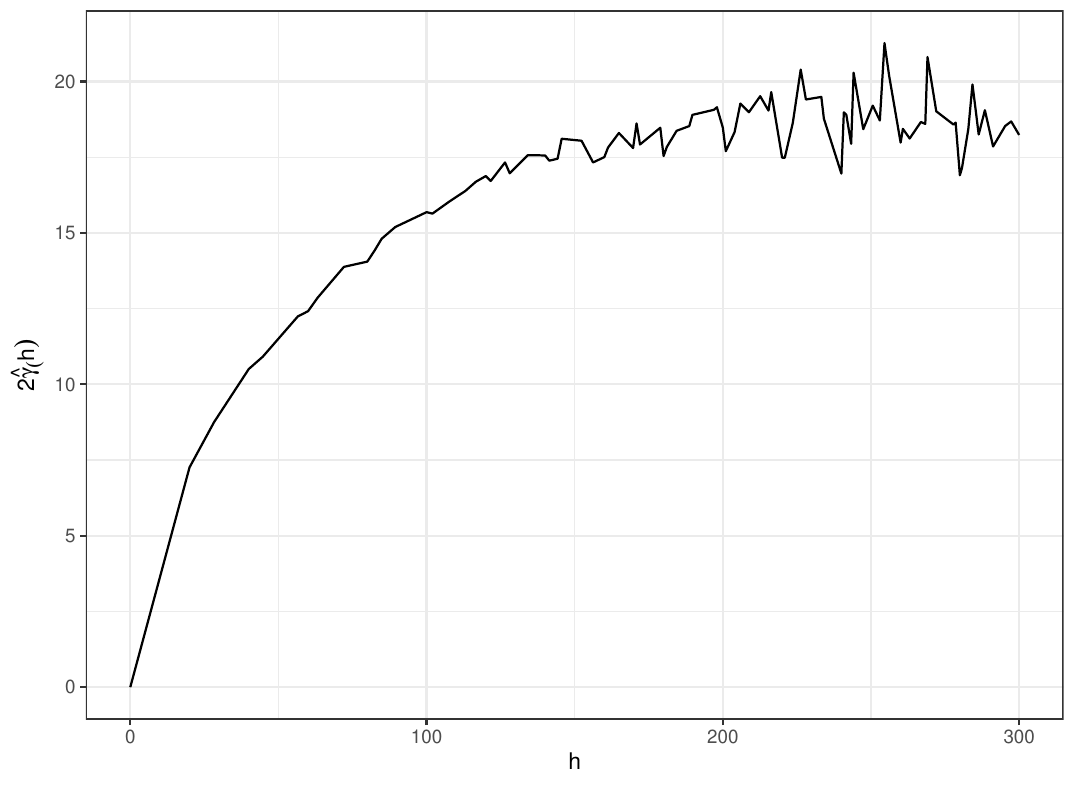}
    \caption{Trace-variogram for smoothed Hill number profiles obtained as in Eq. (\ref{trace_variogram:HP}).
     }  \label{trace:var:Hill}
\end{figure}

Figure %s \ref{trace:var:abu} and 
\ref{trace:var:Hill} shows the (omni-directional) empirical trace-variogram as a function of separation distance $h$ and for $Q=5$. Each point on these plots thus represents an average over a number of pairs of estimated biodiversity profiles that are the same distance apart. The trace-variogram for the smoothed Hill profiles shows the typical increasing trend and reaches an upper bound after the initial increase. This suggests that nearby biodiversity profiles are more correlated and exhibit similar values and so it appears highly informative in the definition of the clusters.

\section{Model-based clustering for spatial functional data}\label{sec5:GMM}

By using the vector of coefficients $\bm{\beta} = (\xi_{0}, \xi_{1}, \alpha_{1}, \ldots, \alpha_{J} )^T$ as representative data for a biodiversity profile, we propose a finite Gaussian Mixture Model (GMM) with a $L_1$ penalized likelihood for functional clustering, named \textit{Penalized model-based Functional Clustering} (PFC-$L_1$) in \cite{pronello2022}.
If a latent variable $Z_i=\{Z_{i1},...,Z_{iK}\}$ denotes the cluster membership of the $i$-th curve to the $k$-th group, the marginal density of $\bm{\beta}$ is a weighted combination of $K$ (number of groups) Gaussian densities $f_k$ with mean vector $\boldsymbol{\mu}_k$ and covariance matrix $\boldsymbol{\Sigma}_k$, that is
$$f(\bm{\beta})=\sum_{k=1}^{K}\pi_{k}(\bm{v};\bm{\omega}) f_k(\bm{\beta};\boldsymbol{\mu}_k, \boldsymbol{\Sigma}_k),$$ 
where $\pi_{k}(\bm{v};\bm{\omega})$ are spatially varying mixing proportions (changing with the spatial coordinate $(x,y)$ of the cell $\bm{v}$ and such that $\sum_{k=1}^K\pi_{k}(\bm{v};\bm{\omega})=1$) depending on some parameters $\bm{\omega}$ that, \textit{a priori}, give the probabilities of belonging to a group, i.e. $\pi_k(\bm{v};\bm{\omega})=\mathbb{P}(Z_k(\bm{v})=1)$, $k=1, \ldots, K$, and $\pi_k(\bm{v};\bm{\omega})>0$ for each $k$.
Then we can write the log-likelihood function as
$$l(\boldsymbol{\theta};\boldsymbol{\beta})=\sum_{i=1}^N log \left[ \sum_{k=1}^K \pi_{k}(\bm{v};\bm{\omega}) f_k(\bm{\beta}_i;\mu_k,\Sigma_k) \right],$$
where $\boldsymbol{\theta}$ 
is the set of all model parameters to be estimated, while 
$\bm{\beta}_i = (\xi_{0i}, \xi_{1i}, \alpha_{1i}, \ldots, \alpha_{Ji} )^T$ is the vector of coefficients of the $i$-th biodiversity profile. 

\subsection{Spatial modelling of mixing proportions}\label{sec:spatial}
Spatially varying mixing proportions are introduced in the GMM model to take into account the spatial dependence among biodiversity profiles. We thus assume that observations corresponding to nearby locations are more likely to have similar allocation probabilities than observations that are far apart in space.

Considering the $K$-th group as a baseline, let
\begin{equation}
\zeta_k(\bm{v};\bm{\omega}) = \log \bigl(\pi_k(\bm{v};\bm{\omega}) / \pi_K(\bm{v};\bm{\omega})\bigr), \qquad k=1, \ldots, K-1,
\label{odds}
\end{equation}
denote the log-odds spatial process. Also, let  $\bm{V}$ be a valid $(N \times N)$ \textit{generalized} variogram matrix \citep{ChilesDelfiner} and $\bm{U}$ a $(N \times 3)$ design matrix whose rows are defined as $\bm{u}_i=(1, x_{i}, y_{i})^{T}$, where $(x_i,y_i)$ are the spatial coordinates of the cell $\bm{v}_i$. Then, if we define the so called \textit{Bending Energy} matrix \citep{Mardia1998} as
$$\mathbf{B}=\mathbf{V}^{-1}-\mathbf{V}^{-1}\mathbf{U}\left(\mathbf{U}'\mathbf{V}^{-1}\mathbf{U}\right)^{-1}\mathbf{U}'\mathbf{V}^{-1},$$
it can be shown - as a result of the Karhunen-Lo\'eve (KL) theorem \citep{Adler2010} - that the log-odds spatial process $\zeta_k(\bm{v};\bm{\omega})$ can be rewritten as a linear model through the following truncated KL expansion
\begin{equation}\label{eq:multinomial_logit2}
    \zeta_k(\bm{v_i};\bm{\omega})= 
 \sum_{l=1}^L \omega_{l,k} \  \psi_{l}(\bm{v}_i), \quad i=1, \ldots, N,   
\end{equation}
where $\omega_{l,k}$ are the elements of the vector $\bm{\omega}$ to be estimated, and the $\psi_{l}(\bm{v}_i)$ are basis functions defined as the eigenvectors obtained by the spectral decomposition  $\mathbf{B}=\bm{\Psi} \mathbf{G} \bm{\Psi}'$, with $\mathbf{G}=diag(g_1, \ldots, g_N)$ being the diagonal matrix of eigenvalues. Since it can be shown that $\mathbf{B}\mathbf{U}=\bm{0}$, it follows that the first three eigenvalues of $\mathbf{B}$ are equal to zero and the corresponding eigenvectors are given by the columns of $\mathbf{U}$. 

In practice, the modelling of the log-odds spatial process is facilitated by the truncated KL expansion based on the property that, given any orthonormal basis functions,
we can find some integer $L$ so that $ \zeta_k(\bm{v};\bm{\omega})$ can be approximated by the finite weighted sum of basis functions. It can be shown \citep{mardia1996kriging} that, when the variogram matrix is parametrized as follows $$V(h_{i,r})=\frac{1}{8\pi}h_{i,r}^2\log(h_{i,r}),$$
where $h_{i,r}=||\bm{v}_i-\bm{v}_r||_2$ and the basis functions $\psi_{l}(\bm{v}_i)$ are obtained through the spectral decomposition of $\mathbf{B}$ above, the spatial process $ \zeta_k(\bm{v};\bm{\omega})$ is modelled through a \textit{Thin-plate spline}. 

\subsection{Penalized likelihood}

Allowing for different cluster means and covariance matrices the specified model can be over-parametrized, and to keep flexibility we avoid introducing any kind of constraints by, instead, considering two penalties that regularize parameter estimation in the log-likelihood function, as in \cite{ZhouETal2009}. Thus, given the profile coefficients $\boldsymbol{\beta}_i$ with length $p=J+2$, and conditional on the number of groups $K$, the  penalized log-likelihood function can be written as
\begin{equation}\label{eq:Lik1}
l_P(\boldsymbol{\theta};\boldsymbol{\beta})=\sum_{i=1}^{N}log \left[\sum_{k=1}^{K}\pi_k(\bm{v};\bm{\omega}) f_k(\boldsymbol{\beta}_i;\boldsymbol{\mu}_k,\boldsymbol{\Sigma}_k)\right]-\lambda_1\sum_{k=1}^{K} \sum_{j=1}^{p}|\mu_{k,j}|-\lambda_{2}\sum_{k=1}^{K}\sum_{j,q}^{p}|W_{k;j,q}|,
\end{equation}
where $\lambda_1>0$ and $\lambda_2>0$ are tuning parameters to be suitably chosen, $\mu_{k,j}$ are cluster mean elements and $W_{k;j,q}$ are entries of the inverse of the cluster-specific covariance matrix
$\mathbf{W}_k=\boldsymbol{\Sigma}_{k}^{-1}$. 
The name \textit{Penalized model-based Functional Clustering} (PFC-L$_1$) in \cite{pronello2022} is chosen because the penalty terms contain sums of absolute values, and so they are of $L_1$ (or LASSO) type. Indeed, the first penalty term facilitates the selection of basis functions appearing in the expansion of $\tilde{H}$ by keeping only the terms useful in separating groups. The second penalty term helps to shrink the elements $W_{k;j,q}$ and allows estimating - thanks to sparsity - large covariance matrices and avoiding possible singularity problems.

The model parameter estimation cannot be obtained by direct optimization of the log-likelihood function given in Eq.~(\ref{eq:Lik1}) but, since $Z$ is not observed, can be efficiently carried out using the Expectation-Maximization (EM) algorithm \citep{EM}. The analytical solutions to update the cluster membership probabilities, the cluster mean elements and the cluster-specific precision matrices are detailed in \cite{pronello2022}. In particular, at each iteration the Graphical LASSO algorithm \citep{gLASSO} is used to obtain sparse cluster-specific precision matrices, whereas to estimate the spatially varying mixing proportions $\pi_{k}(\bm{v};\bm{\omega})$ the multinomial logit model as specified in Section \ref{sec:spatial} needs to be fitted.
Thus, the estimation of the parameters of the linear model in Eq.~(\ref{eq:multinomial_logit2})
can be obtained at the $(d+1)$-th iteration 
of the EM algorithm as the solution of the log-likelihood maximization of a weighted multinomial logit model, that is 
$$ \widehat{\bm{\omega}}^{(d+1)} = \underset{\bm{\omega}}{arg \ max} \sum_{i=1}^N \sum_{k=1}^K \widehat{\tau}_{k}^{(d)}(\bm{v}_i) \log \bigl(\pi_k(\bm{v}_i;\bm{\omega})\bigr),$$
where $\hat{\tau}_{k}^{(d)}(\bm{v}_i)$ are the estimated posterior probabilities that a biodiversity profile $i$, summarized here by $\hat{\bm{\beta}}_i$, belongs to the $k$-th group, and are computed through the iterations of the EM algorithm as 
\begin{equation}\label{eq:post_prob}
\widehat{\tau}_{k}^{(d)}(\bm{v}_i)=\frac{\widehat{\pi}_{k}^{(d)}(\bm{v}_i;\bm{\omega})
f_k(\hat{\bm{\beta}}_i;\widehat{\boldsymbol{\mu}}_k^{(d)},\widehat{\boldsymbol{\Sigma}}_k^{(d)})}{\sum_{k=1}^{K} \widehat{\pi}_{k}^{(d)}(\bm{v}_i;\bm{\omega})f_k(\hat{\bm{\beta}}_i;\widehat{\boldsymbol{\mu}}_k^{(d)},\widehat{\boldsymbol{\Sigma}}_k^{(d)})}.
\end{equation}

\subsection{Model selection}

One of the most difficult steps in clustering is to determine the optimal number of clusters, $K$, to group the data, and we know there is no "right" answer. In this paper, we perform a grid-search for model hyper-parameters and choose the triplet $\{K;\lambda_1;\lambda_2\}$ that allows for model selection based on information criteria. In particular, we consider likelihood-based measures of model fit that include a penalty for model complexity such as the Bayesian Information Criterion (BIC) 
$$BIC(K, \lambda_1, \lambda_2)= l(\hat{\boldsymbol{\theta}}_K;\hat{\boldsymbol{\beta}} \mid K, \lambda_1, \lambda_2) -\frac{C}{2}\log(N)$$
and the Integrated Classification Likelihood (ICL) index \citep{Baudry2015} 
$$ICL(K, \lambda_1, \lambda_2)= BIC(K, \lambda_1, \lambda_2) + \sum_{k=1}^K\sum_{i=1}^N 
\hat{\tau}_{k}(\bm{v}_i) \log \hat{\tau}_{k}(\bm{v}_i)$$
where $l(\hat{\boldsymbol{\theta}}_K;\hat{\boldsymbol{\beta}} \mid K, \lambda_1, \lambda_2)$ is the value of the maximized log-likelihood objective function with parameters $\hat{\boldsymbol{\theta}}_K$ estimated under the assumption of a model with $K$ components, $\hat{\bm{\beta}}$ collects all $\hat{\bm{\beta}}_i$ and $C$ measures the complexity of the model. 
While BIC has a penalty term only related to the number of observations $N$ and the complexity measure $C$, ICL also includes an additional term - that is the estimated mean entropy - to penalize clustering configurations with overlapping groups (this facilitates solutions with well-separated groups, i.e. with low entropy).

To use the above criteria it is necessary to clarify what is $C$ in a penalized model. In our case, we consider
$$ C= \sum_{k=1}^{K} \sum_{j=1}^{p} I \Bigl(\hat{\mu}_{k,j} \neq 0\Bigr)  + \sum_{k=1}^K \sum_{i \leq j} I\Bigl( \widehat{{\Sigma}}_{k;j,q} \neq 0\Bigr) + L(K-1)$$
where $I(\cdot)$ is the indicator function that applies to the (sparse) likelihood estimate of $\boldsymbol{\mu}_{k}$ and $\boldsymbol{\Sigma}_k$, so that $C$ is the number of nonzero entries in both the means and the upper half of the covariance matrices, plus the number of parameters for the spatial mixing proportions. In general, the model with the highest values of BIC or ICL could be selected as the desired model.

\section{Results}\label{sec6:results}

\begin{figure}
\centering
\includegraphics[scale = 0.8]{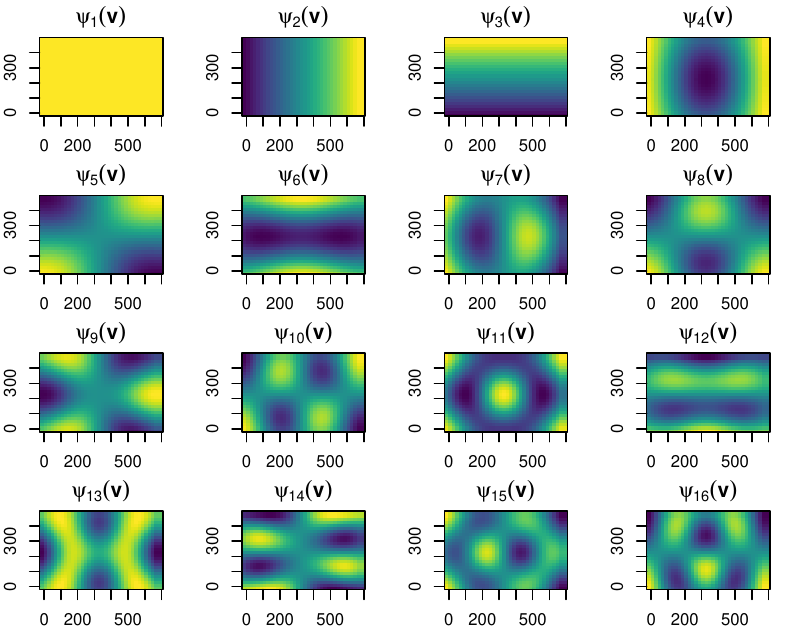}
\caption{Spatial maps of the first 16 basis function $\psi_l$, $l = 1, \ldots, L$, obtained by the spectral decomposition of the \textit{Bending Energy} matrix and used to model the spatial variability of the log-odds as in Eq. (\ref{odds}).}
\label{basis_functions}
\end{figure}

In this section, we extend the statistical analysis of the dataset discussed in Section \ref{sec2:data} and present the results obtained from clustering the biodiversity profiles using the PFC-$L_1$ procedure. The analyses are carried out by developing custom code within the \texttt{R} environment \citep{R}. To take care of the spatial dependence among the profiles, we have considered a Thin-plate spline parametrization (see Section \ref{sec:spatial}) with  $L = 16 << N$ basis functions explaining about $91.50\%$ of the spatial variability. The spatial patterns of the basis functions are shown in Figure \ref{basis_functions} and, as expected, they show a decreasing order of smoothness. For example, the first basis function $\psi_1(\bm{v})$ is constant over all the domain of interest while $\psi_2(\bm{v})$ and $\psi_3(\bm{v})$ are linear trends of the longitude and latitude coordinates, respectively. More in general, higher order functions correspond to larger-scale features while lower-order functions correspond to smaller-scale details. 

\begin{figure}
\centering
\includegraphics[scale = 0.6]{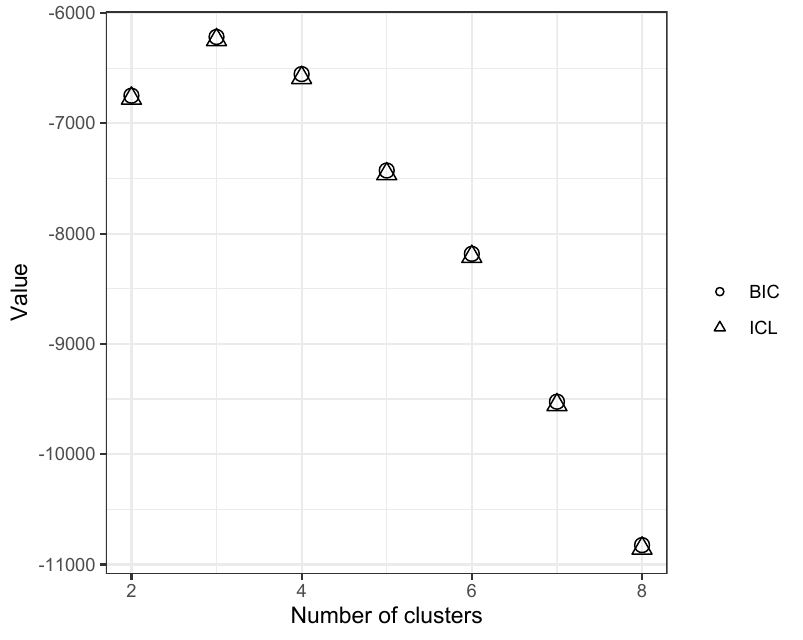}
\caption{BIC and ICL values for model selection. The plot maps the maximum BIC and ICL values achieved for the triplet $(K,\lambda_1,\lambda_2)$ according to the number of  clusters $K$.} \label{BIC_ICL}
\end{figure}

By fixing $J = 15$ in Eq.~\eqref{eq:fit_fuct} and considering a discrete grid of values for the triplet $(K, \lambda_1,\lambda_2)$, the BIC and ICL criteria suggest that a GMM model with three spatial clusters should be considered (see Figure \ref{BIC_ICL}). BIC and ICL values closely align since the posterior probability estimates result in distinct partitions, where the clusters are well-separated with estimated mean entropy approaching zero. However, we are not aware of the original distribution which generated the data so, to validate the performance evaluation of the clustering process we also consider interpretation as an important part of model selection, especially from a knowledge discovery perspective. Interpretation can help us gain insights and guiding decisions based on our clustering procedure and for this, in the following, we favour the solution with $K = 4$ as it better highlights the group of cells with constant biodiversity profiles (see below) and for which the values of BIC and ICL are the “second best”.

\begin{figure}\centering
\subfloat{\hspace{-0.5cm}\includegraphics[height = 4cm, width = 5cm]{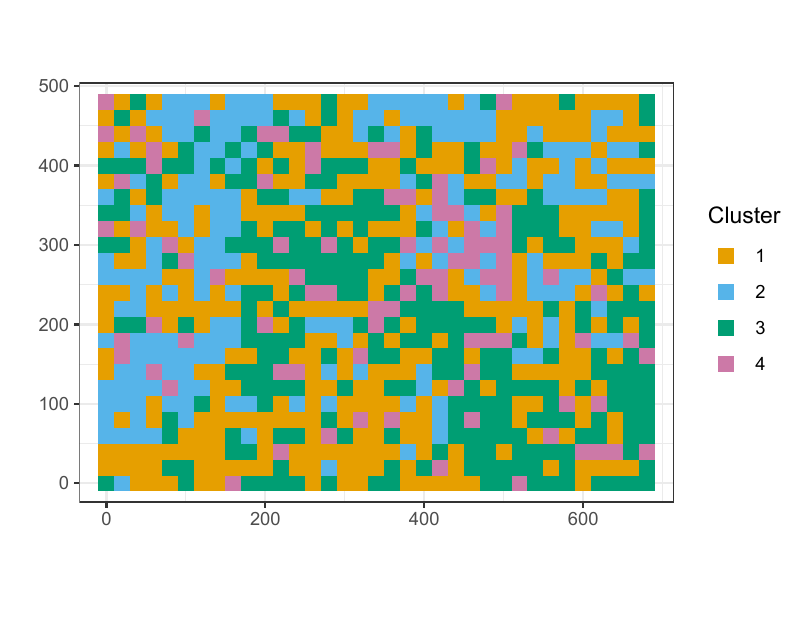}} \hspace{0.5cm}
\subfloat
{\includegraphics[height = 4cm, width = 6cm]{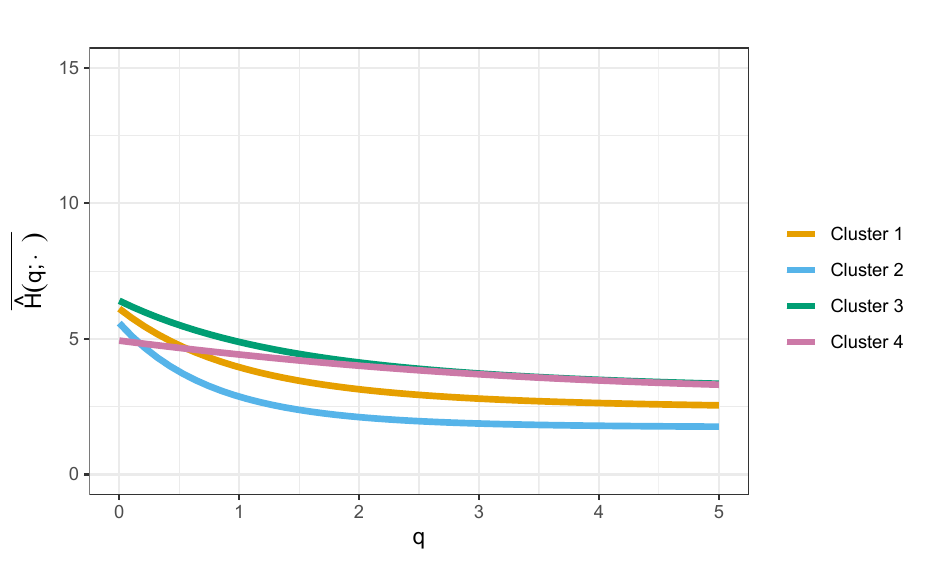}}\par 
\hspace{-2cm} \subfloat{\includegraphics[scale = 0.70]{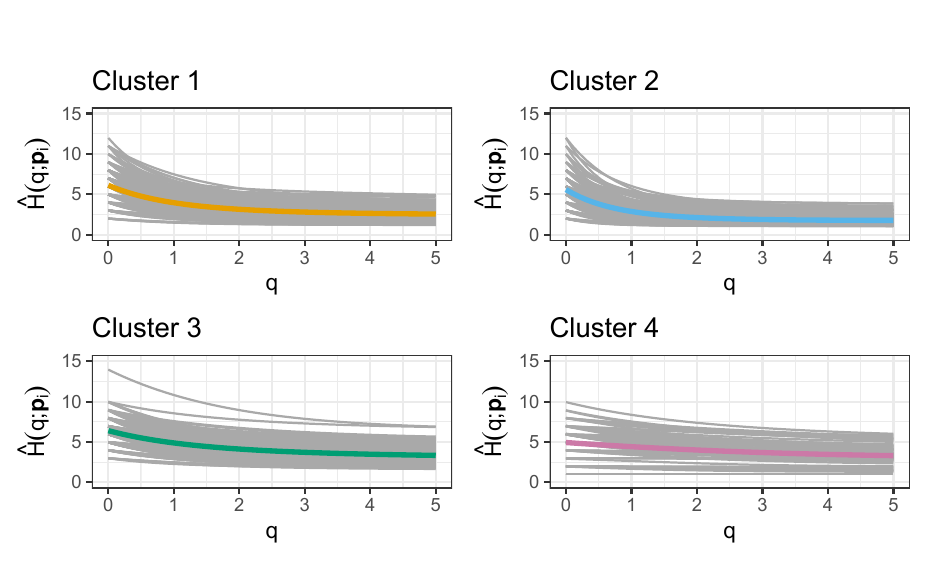}}
\caption{\textit{Upper left}: Functional zoning results of the Prospect Hill Tract long-term plot with four clusters (each cell is assigned a specific colour based on its associated clustering label). \textit{Upper right}: estimated mean biodiversity profiles $\overline{\hat{H}(q;\cdot)}$ in each cluster. \textit{Bottom}: individual biodiversity profiles $\hat{H}(q;\bm{p}_i)$ in each cluster with superimposed estimated mean biodiversity profiles (thicker lines).}
\label{zoning}
\end{figure}

Figure \ref{zoning} provides a spatial representation of the four clusters. In particular, the upper left panel illustrates the functional zoning of the Prospect Hill Tract long-term plot derived from these clusters, the upper right panel displays the behaviour of the estimated mean biodiversity profiles and the bottom panel exhibits the allocation of the individual biodiversity profiles $\hat{H}(q;\bm{p}_i)$ in each cluster. Due to the intersection of the estimated mean biodiversity profiles, direct comparisons among the four clusters are not feasible, as the profiles only offer a partial ordering of their diversities. Although this limitation cannot be entirely overcome, biodiversity profiles remain significantly more meaningful than univariate indices. 
In fact, even in cases where two communities (cells) are not directly comparable, examining where their biodiversity profiles intersect can reveal changes or variations in the composition of species.

\begin{figure}
\centering
    \includegraphics[scale = 0.5]{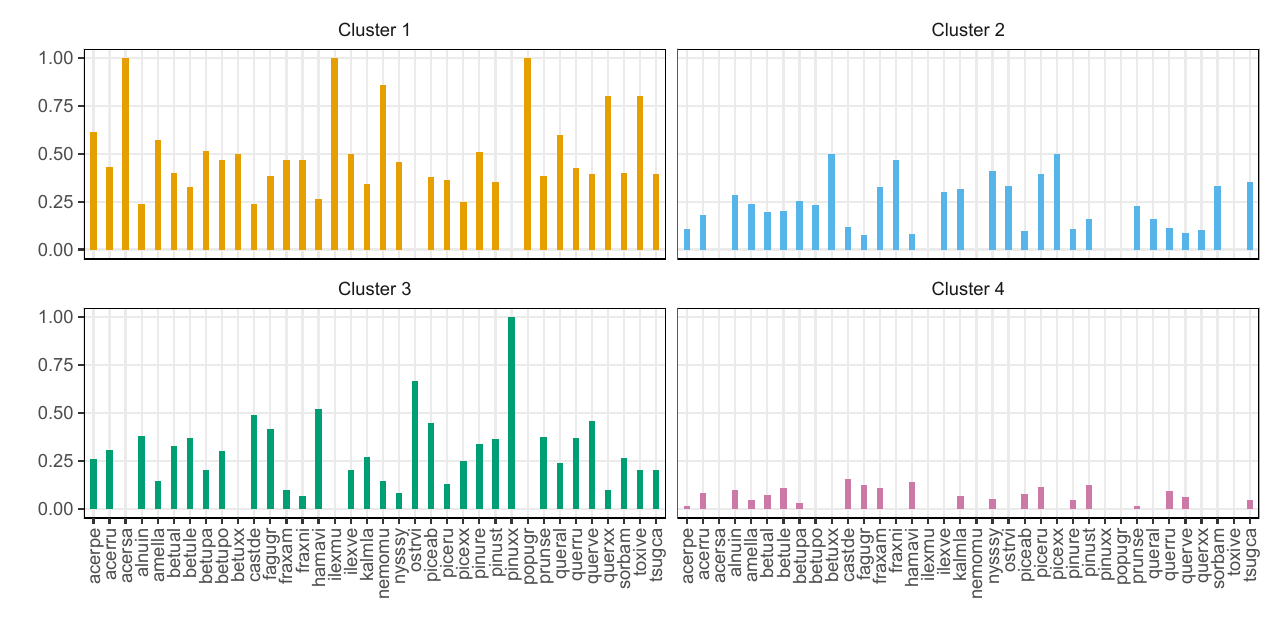}
    \caption{Distribution of species in each cluster.}
    \label{distr_species}
\end{figure}

Cluster 1 and Cluster 3 emerge as the most populated clusters, with 326 and 264 cells, respectively, whereas Cluster 2 includes 196 cells and, finally, Cluster 4 only contains 89 cells. All clusters display similar average species richness (when $q = 0$) despite different levels of variability and slope, as shown in the bottom panel of Figure \ref{zoning}. In particular, Cluster 4 exhibits the lowest average species richness among the clusters. Remarkably, the clusters exhibit diverse species compositions, implying that they achieve similar average species richness by having unique sets of species in each cluster. For example, Cluster 1 includes solely one \textit{Acer saccharum} tree, while this particular species is entirely absent in Cluster 3 as illustrated in Figure \ref{distr_species}. 

Although all clusters have similar average species richness, they show different values for average species abundance (when $q = 1$) and average species dominance (when $q = 2$). For example, compared with Clusters 1 and 2, Cluster 4 displays higher average species abundance and dominance resulting from estimated mean biodiversity profile intersections. These findings emphasize the nuanced differences in species distribution and dominance within the identified clusters. The upper right plot of Figure \ref{zoning} further confirms that for $0 \leq q \leq 2$, the biodiversity profiles are sufficient to characterize the \textit{taxonomy diversity} in the Prospect Hill Tract long-term plot.

In general, the main contributing factor in differentiating between the clusters appears to be associated with the derivatives of the estimated Hill profiles. These derivative functions convey significant information and are consistent with the functional representation used in Eq. (\ref{eq:fit_fuct}). Clusters 1 and 2 are characterized by curves with steeper slopes, while Cluster 4 stands out with profiles that remain relatively constant regardless of the intercept level. This behaviour holds particular significance when interpreting the clustering results since, as demonstrated in the example from Section \ref{sec3:biodiv}, a constant profile indicates a uniform distribution of species within the cell, while a more convex profile suggests an uneven distribution.

\begin{figure}
     \centering
         \centering
         \includegraphics[scale = 0.7]{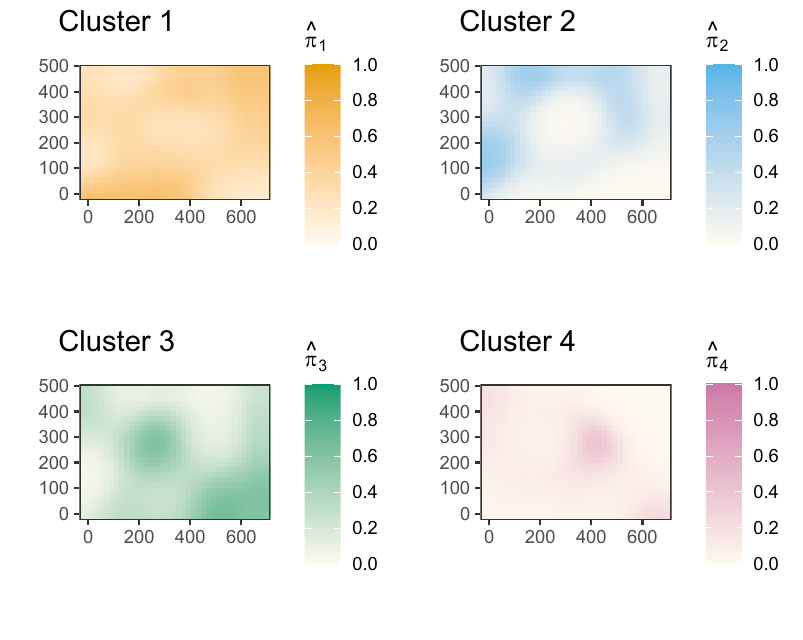}
     \caption{Maps of the estimated prior probabilities $\hat{\pi}_k(\bm{v};\bm{\omega})$ for each cluster of the Prospect Hill Tract long-term plot.}     
    \label{prior}
\end{figure}

Figure \ref{prior} displays the spatial distribution of the estimated prior probabilities $\hat{\pi}_k(\bm{v};\bm{\omega})$ for each cluster. As it can be noticed, the distribution of the clusters clearly shows how the estimated posterior probabilities, $\hat{\tau}_k(\bm{v}_i)$, reflect the information about the spatial distribution of the weights of the mixture (see upper left panel Figure \ref{zoning}). As illustrated in Section \ref{sec5:GMM}, we note that clusters arise from a careful balance between geographical proximity and similarity among curves (biodiversity profiles). The values represented by $\hat{\pi}_k(\bm{v};\bm{\omega})$ provide valuable information about the spatial variability of the clusters. Consequently, the outcomes shown in Figure \ref{prior} serve as spatial predictions of the clustering labels, focusing solely on spatial information. These predictions enable us to divide the study area into distinct zones that highlight the prevalence of specific clusters, offering policymakers insightful guidance for crafting effective interventions. For instance, policymakers could establish appropriate perimeters for areas at risk based on the clustering results and estimated prior probability maps, optimizing their decision-making process and resource allocation.

\section{Discussion}\label{sec7:discussion}

Biodiversity profiles present a valuable tool for researchers to characterize and compare ecological communities by accounting for both abundant and rare species, thus recognizing the multidimensional aspects of diversity. In this study, following \cite{gattone2009functional}, we have treated biodiversity profiles as non-negative and convex curves, amenable to analysis through FDA methodologies. In particular, by considering the whole profiles as single entities, we have integrated functional data analysis with spatial (model-based) clustering techniques to identify and delineate homogeneous zones based on spatial contiguity and shape similarity of the curves. This approach goes beyond traditional methods that may consider only individual abundance vectors and offers a more comprehensive understanding of biodiversity distribution, capturing the underlying patterns and variations across different regions. By focusing our study on a plot of the Harvard Forest, classification results indicate that our modelling approach can provide valuable information for policymakers, enabling them to make informed decisions regarding the conservation and management of natural resources. 

However, due to the lack of additional information in the available data, we acknowledge a few limitations in our taxonomic diversity. For example, all species are treated as equally distinct from one another, disregarding potential species differences in our study. In general, biodiversity extends beyond mere species diversity, encompassing a broader spectrum that includes phylogenetic, genetic, and functional diversity \citep{Pielou1975}. Relying solely on species names provides limited insights into the functions or evolutionary history of these species, which are instead crucial for understanding the underlying processes contributing to the observed levels of biodiversity. However, despite the acknowledged limitations, there are promising avenues to enhance our functional framework for biodiversity profiles. One approach involves incorporating pairwise similarities between species using a similarity matrix, leading to the \textit{Leinster-Cobbold diversity} of order $q$ as proposed by \cite{Leinster2012}. Alternatively, we can explore the unified framework proposed in  \cite{ChaoColwell2022}, which defines the \textit{Hill-Chao numbers} of order $q$ to assess biodiversity across multiple dimensions. By incorporating species trait similarities or adopting the more general framework of \cite{ChaoColwell2022}, we can gain a more complete understanding of a community and improve predictions of ecosystem functions. These approaches represent promising directions for future research, aiming to provide a more nuanced and comprehensive perspective of biodiversity dynamics and their ecological significance.

%\section*{Acknowledgments}

%\subsection*{Author contributions}

%\appendix

%\section{Section title of first appendix\label{app1}}

%\nocite{*}% Show all bib entries - both cited and uncited; comment this line to view only cited bib entries;
\bibliographystyle{apalike}
\bibliography{biblio}

%\section*{Author Biography}
%\begin{biography}{\includegraphics[width=60pt,height=70pt,draft]{empty}}{\textbf{Author Name.} This is sample author biography text.}
%\end{biography}

\end{document}